# Evidence of coherent c-axis charge transport in underdoped La$_{2-x}$Sr$_x$CuO$_4$ superconductors

Y. H. Kim,[1] P. H. Hor,[2] X. L. Dong,[3,4] F. Zhou,[4] Z. X. Zhao,[4] Z. Wu,[2] and J. W. Xiong[4]

[1]*Department of Physics, University of Cincinnati, Cincinnati, Ohio 45221-0011, USA*
[2]*Department of Physics and Texas Center for Superconductivity, University of Houston, Houston, Texas 77204-5005, USA*
[3]*Texas Center for Superconductivity, University of Houston, Houston, Texas 77204-5002, USA*
[4]*National Laboratory for Superconductivity, Institute of Physics and Center for Condensed Matter Physics, Chinese Academy of Sciences, P.O. Box 603, Beijing 100080, People's Republic of China*

We have observed the plasma edge in the *normal state* c-axis far-infrared (far-IR) reflectivity of La$_{2-x}$Sr$_x$CuO$_4$ (LSCO) single crystals at an unlikely Sr-doping concentration x=0.07 but not in x=0.09. We find that the c-axis scattering rate ($\Gamma_c$) is surprisingly small, $\Gamma_c = 13 \pm 2$ cm$^{-1}$ for x=0.07 and $\Gamma_c = 24 \pm 1$ cm$^{-1}$ for x=0.09, and nearly temperature-independent but increases linearly with doping up to x=0.16. We suggest that the apparent absence of the normal state plasma edge in the previous measurements is due to the larger $\Gamma_c$ than the screened plasma frequency of the coherent free carriers. We conclude that the c-axis charge transport in LSCO in the underdoped regime is intrinsically *coherent*.

The answer to the fundamental question of whether or not the ground state of the high-temperature superconducting (high-$T_c$) cuprate can be treated as Landau's Fermi liquid hinges critically on the nature of charge dynamics in the direction perpendicular to the CuO$_2$ planes (c axis).[1] There have been many puzzling but experimentally sound observations of the c-axis charge behaviors in the underdoped regime. For instance, the c-axis resistivity ($\rho_c$) measurements, on the one hand, revealed a rather complex behavior with the doping concentration of the holes (p).[2] In the underdoped regime, the c-axis transport appears incoherent as if the phase of the carriers is completely random and the wave vector is not conserved but becomes coherent at the optimal doping concentration (p=0.16) and beyond.[3] The Fermi liquid ground state of the overdoped Tl$_2$Ba$_2$CuO$_{6+\delta}$ has been confirmed by the recent observation of the c-axis Fermi surface.[4] However, the interpretation of the above results as a simple crossover from non-Fermi-liquid to Fermi-liquid at the optimal doping is not feasible because it is clearly shown that the superconductivity in both underdoped and overdoped La$_{2-x}$Sr$_x$CuO$_4$ (LSCO) exhibits the identical critical behavior under magnetic field,[5] a strong indication of a common mechanism of superconductivity in both regimes. Furthermore, the conflicting observations, where a quasiparticle tunneling behavior is clearly seen in thick (22 $\mu$m) crystals[6] versus a "normal" transport in thin (1–3 $\mu$m) crystals,[7] in two independent but careful measurements on optimally doped and slightly underdoped high-quality Bi-2212 single crystals along the c axis, seem to have no simple answer.

On the other hand, the corresponding c-axis far-infrared (far-IR) reflectivity studies done mostly on the slightly underdoped or optimally doped LSCO seem to agree with the non-Fermi-liquid interpretation due to the absence of the plasma edge in the normal state c-axis reflectivity even at the optimal doping.[8–12] It is found that only in the superconducting state does there develop a plasma edgelike structure out of the otherwise insulatorlike normal state c-axis reflectivity.

However, in their analysis of the c-axis far-IR reflectivity data, the normal state c-axis reflectivities could be unambiguously fitted with the Drude model with $\Gamma_c \sim 120$ cm$^{-1}$ for x=0.13 (Ref. 12) and $\Gamma_c \sim 170$ cm$^{-1}$ for x=0.16.[8] These $\Gamma_c$'s are unexpectedly small considering the surmise that the transport behavior of the free carriers is incoherent, warranting no normal state plasma edge in the reflectivity. Moreover, the increasing tendency of $\Gamma_c$ with doping is highly anomalous and counterintuitive since the c-axis resistivity of x =0.13 LSCO is higher than that of optimally doped x=0.16 LSCO at all temperatures.[3]

In an effort to establish a working charge model that unifies these problematic but robust "experimental facts," we considered the following two scenarios: (1) The intrinsic normal state c-axis charge transport is indeed incoherent and attributable to the non-Fermi-liquid nature of the charge system, and (2) it might be intrinsically coherent, but for some profound reason the measured c-axis conductivity appears to come from incoherent hopping of the carriers. Motivated by the clear observation of the plasma edge in the normal state far-IR reflectivity of the cation and anion codoped polycrystalline LSCO samples with the hole concentration p=0.063 and 0.07 with no appreciable changes in the plasma edge when superconducting,[13] we pursued the second scenario by carrying out the resistivity, thermopower, and far-IR reflectivity measurements along the c axis of the LSCO crystals near the insulator-superconductor phase boundary, p=0.06.

For the measurement of the c-axis reflectivity, we prepared two high-quality LSCO single crystals grown by the traveling-solvent floating-zone method with nominal carrier concentration x=0.07 ($T_c \sim 20$ K) and 0.09 ($T_c \sim 28$ K), which gives the corresponding p=0.07 and 0.09, respectively.[14] The sample area was 5 mm in diameter and the angular error from the desired crystal axis was less than ±1°. We measured the reflectivity at a near normal angle of incidence ($\sim 8°$) on the sample with the far-IR polarization along the c-axis direction. A Bruker 113v spectrometer was used to cover the frequency ($\omega$) between

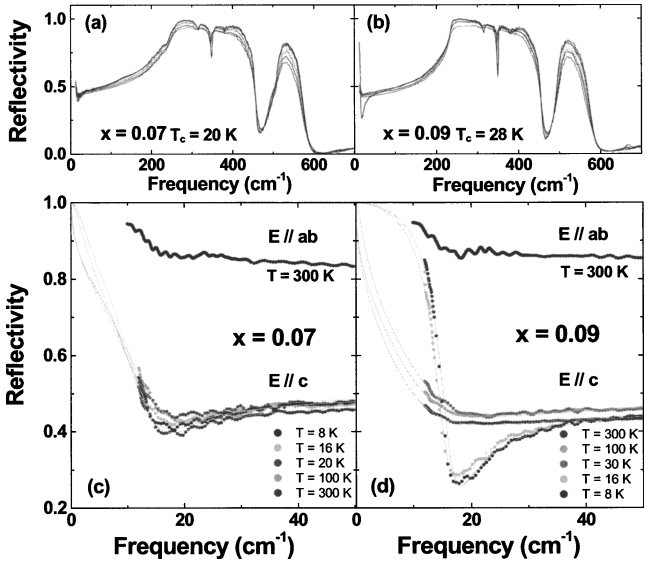

FIG. 1. (Color online) Far-infrared reflectivity spectra of La$_{2-x}$Sr$_x$CuO$_4$ at various $T$ with the far-IR polarization parallel to the $c$ axis (E//$c$). (a) x=0.07 ($T_c \sim 20$ K) and (b) x=0.09 ($T_c \sim 28$ K). (c) and (d) The $c$-axis reflectivity and the corresponding $ab$-plane reflectivity (E//$ab$) shown in an expanded scale.

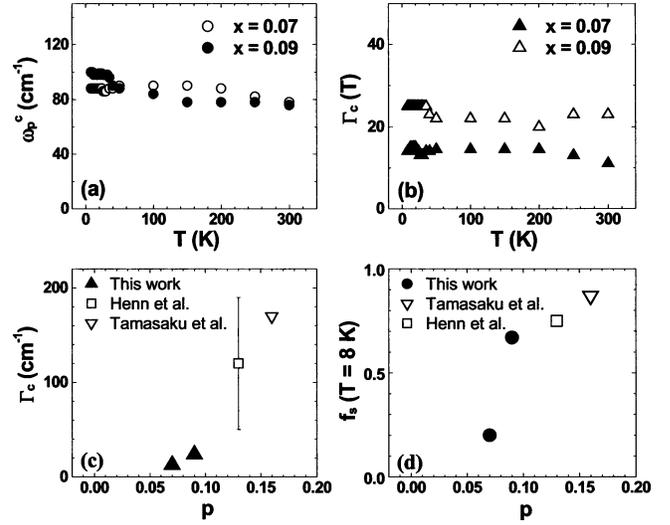

FIG. 2. The $T$ and p dependences of $\omega_p^c$, $\Gamma_c$, and $f_c$ (see the text for definitions). In the fitting process, $\varepsilon_\infty^c = 30 \pm 1$ and $29 \pm 2$ were used for x=0.07 and 0.09, respectively. The open squares in (c) and (d) are taken from Ref. 12 and the open inverse triangles from Ref. 8.

12 cm$^{-1}$ (~1.5 meV) and 4000 cm$^{-1}$ (0.5 eV). As a reference, we used a gold (Au) mirror made by depositing Au film on a stainless steel mirror that had been polished under the same condition as the single-crystal samples. The far-IR beam size was kept at 3 mm in diameter, smaller than the size of the sample (5 mm). We controlled the sample temperature ($T$) by directly monitoring $T$ from the backside of the sample. We extracted the far-IR properties, represented by the real part of the $c$-axis conductivity $\sigma_1^c(\omega)$ and the real part of the $c$-axis dielectric function $\varepsilon_1^c(\omega)$, from a Kramers-Kronig transformation of the reflectivity. For high $\omega$ extrapolation above 4000 cm$^{-1}$ (0.5 eV), we used the published data for LSCO single crystals.[15]

In Fig. 1, we display the $c$-axis far-IR reflectivities of x=0.07 and x=0.09 LSCO measured at various $T$. Overall characteristics of the reflectivity are similar to those previously reported for x=0.13 (Ref. 12) and x=0.16 (Ref. 8) LSCO. However, the low $\omega$ region of the normal state reflectivity, shown in an expanded scale in Figs. 1(c) and 1(d), clearly shows a sharp plasma edge at ~15 cm$^{-1}$ of the x=0.07 sample at all $T$ and a less pronounced but still noticeable broad local minimum at a similar $\omega$ for the x=0.09 sample for $T > T_c \sim 28$ K. No such plasma edge has been observed in the $c$-axis normal state reflectivity of LSCO (0.1 $\leq$ x $\leq$ 0.16). In the superconducting state, while the reflectivity minimum gets dramatically deeper at the same location for the x=0.09 sample, the reflectivity minimum grows only slightly deeper at the same $\omega$ for the x=0.07 sample. It is remarkable to note that this $c$-axis screened plasma edge occurs nearly at the same $\omega$ as that of the $ab$ plane reported recently, as plotted in Figs. 1(c) and 1(d).[16] This is also consistent with and can be regarded as an independent confirmation of the earlier results obtained for polycrystalline samples.[13]

An excellent fit was obtained in the vicinity of the plasma edge by using the dielectric function $\varepsilon^c(\omega) = \varepsilon_\infty^c - f_s (\omega_p^c)^2/\omega^2 - (1-f_s)(\omega_p^c)^2/\{\omega(\omega + i\Gamma_c)\}$, where $\omega_p^c$ is the unscreened $c$-axis plasma, $f_s$ is the superfluid fraction, and $\varepsilon_\infty^c$ is the background dielectric constant along the $c$ axis. We found that as displayed in Fig. 2(a), $\omega_p^c$ has a weak $T$ dependence and $\Gamma_c$ is nearly $T$-independent [see Fig. 2(b)] and extremely small, $\Gamma_c = 13 \pm 2$ cm$^{-1}$ for x=0.07 and $\Gamma_c = 24 \pm 1$ cm$^{-1}$ for x=0.09. The extremely small $\Gamma_c$'s for p=0.07 and 0.09 indicate that the $c$-axis charge transport is not only coherent but also nearly dissipation-free at such a low p and, in terms of $\Gamma_c$, x=0.07 LSCO is better metallic than x=0.09 LSCO. As the doping increases, $\Gamma_c$ increases linearly joining the previously found $\Gamma_c \sim 120$ cm$^{-1}$ for x=0.13 and ~170 cm$^{-1}$ for x=0.16 as plotted in Fig. 2(c).

The nearly $T$-independent $\Gamma_c$ that increases linearly with x suggests that the dominant scattering mechanism along the $c$ axis is the Coulomb scattering of the electrons by the Sr counterions residing between the CuO$_2$ layers. Furthermore, such small $\Gamma_c$'s are not consistent with the incoherent transport of the confined holes in the CuO$_2$ planes, suggesting that the $c$-axis charge transport, within the appropriate far-IR length scale of a few microns, has to be bandlike. Moreover, we found that the $c$-axis superfluid fraction is always partial; $f_s$ is only 0.22 for x=0.07 and 0.69 for the x=0.09 sample at $T=8$ K and reaches $f_s \sim 0.75$ for x=0.13 at $T=10$ K (Ref. 12) and $f_s \sim 0.87$ for x=0.16 at $T=8$ K (Ref. 8) ($f_s = \omega_{sp}^2/\omega_{np}^2$ in the notation used in Refs. 8,12) as shown in Fig. 2(d). This partial $f_s$ at $T \ll T_c$ suggests that the phase-coherence development of the Cooper pairs along the $c$ axis is incomplete even at x=0.16.

The calculated $\sigma_1^c(\omega)$ and $\varepsilon_1^c(\omega)$ clearly show the Drude-like coherent charge dynamics; there is a sharp upturn in the tail of $\sigma_1^c(\omega)$ below $\sim 20$ cm$^{-1}$ and the corresponding $\varepsilon_1^c(\omega)$ decreases as depicted in Fig. 3. Using the parameters found from the fit and the expression for the zero crossing frequency $\omega_p^* = \sqrt{(\omega_p^c)^2/\varepsilon_\infty^c - \Gamma_c^2}$, we found that $\omega_p^* \sim 9$ cm$^{-1}$ at $T$=300 K and $\sim 5.6$ cm$^{-1}$ at $T$=22 K for x=0.07. However, for x=0.09, $\varepsilon_1^c(\omega)$ does not cross zero in the normal state because $\Gamma_c$ is larger than the screened plasma frequency $\omega_p^c/\sqrt{\varepsilon_\infty^c}$. Our results suggest that, for x>0.07, we do not expect to observe the normal state plasma edge in the normal state $c$-axis reflectivity even though the $c$-axis transport is coherent because of the increase of $\Gamma_c$ with Sr doping but the superfluid contribution to $\varepsilon_1^c(\omega)$, given by the term $\varepsilon_1^{sc}(\omega) = -f_s(\omega_p^c)^2/\omega^2$, generates a plasma edge when $T<T_c$.

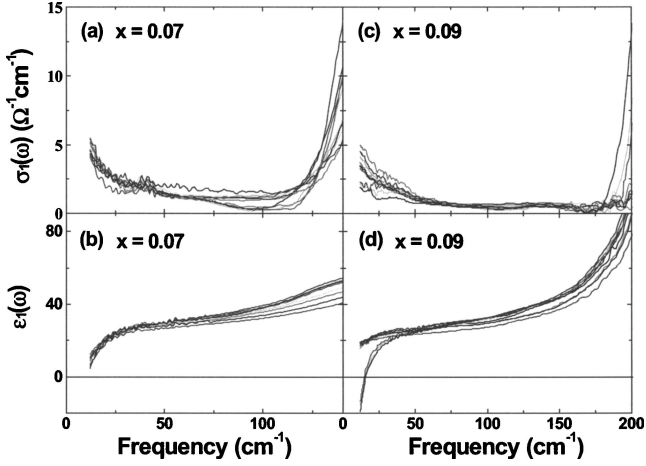

FIG. 3. (Color online) $\sigma_1(\omega)$ and $\varepsilon_1(\omega)$ obtained for x=0.07 and x=0.09 at $T$=300 K, 250 K, 200 K, 150 K, 100 K, 50 K, 30 K, 20 K, 16 K, and 8 K.

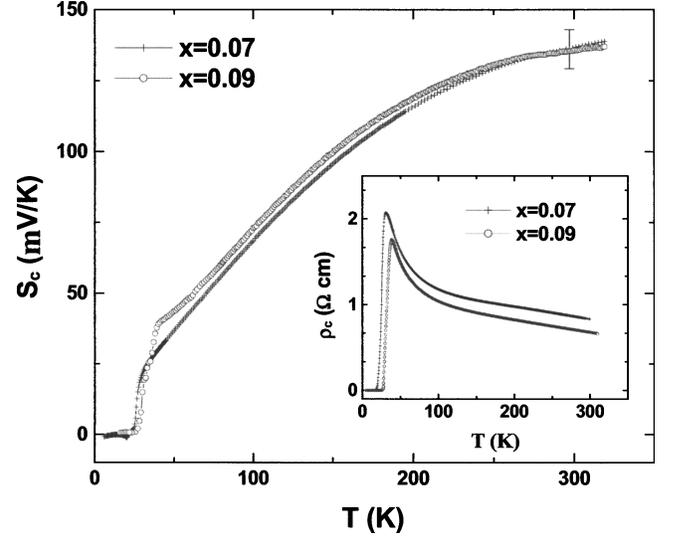

FIG. 4. (color online) The $c$-axis thermopower ($S_c$) for x=0.07 and x=0.9 LSCO as a function of $T$. The error bar for x=0.07 is due to the inherent electronic inhomogeneity in the highly underdoped LSCO. Inset: $T$ dependence of $\rho_c$ of x=0.07 and 0.09 LSCO.

While our far-IR results seem to paint a convincing picture of the coherent nature of the $c$-axis charge carriers, the normal state $\rho_c$ with $d\rho_c/dT<0$ shown in the inset of Fig. 4 suggests otherwise. This can be understood by noting that there is a difference in the nature of the dc and far-IR conductivity measurements: the dc transport measurement is in effect the measurement of the resistors connected in series and, hence, requires a complete metallic path along the $c$ axis in order to obtain a metallic transport while the far-IR probe only needs a metallic domain of its wavelength scale.[17] Therefore, we can have both the incoherent characteristics in $\rho_c$ and the coherent behavior in the $c$-axis far-IR reflectivity if we have stacks of a few-micron-thick "3D metallic slabs" of LSCO where the insulating gaps between the slabs are randomly formed along the $c$ axis by the commonly accepted electronic disorder.[18] This prohibits the global $c$-axis energy band formation in the underdoped regime all the way up to optimal doping.

The validity of this model can be independently tested by carrying out the thermopower measurements because thermopower is sensitive to and dominated by the metallic response. If the proposed model is correct, then we expect to observe a decreasing thermopower with decreasing $T$ even though the resistivity data show the opposite behavior. Furthermore, one should be able to confirm the better metallic behavior along the $c$ axis for x=0.07 ($\Gamma_c=13\pm 2$ cm$^{-1}$) than for x=0.09 ($\Gamma_c=24\pm 1$ cm$^{-1}$) in the thermopower data. The thermopower data displayed in Fig. 4 demonstrate that the $c$-axis thermopower ($S_c$) decreases with decreasing $T$ and x=0.07 LSCO is intrinsically better metallic. Furthermore, this picture seems to clearly and consistently explain why the quasiparticle tunneling is observed in thick (22 $\mu$m) (Ref. 6) and a "normal" transport is observed in thin (1–3 $\mu$m) (Ref. 7) high-quality Bi-2212 single crystals along the $c$ axis.

In summary, we present evidence that the charge transport along the $c$ axis is intrinsically coherent. The incoherent behavior in $\rho_c$ at x<0.16 comes from the hopping of the coherent free carriers across the insulating barriers distributed along the $c$ axis in a macroscopic length scale of a few microns. We have identified the physical origin of the apparent absence of the normal state plasma edge in the $c$-axis far-IR reflectivity despite the recovery of the coherent $c$-axis dc transport at the optimal doping, which is due to a larger $\Gamma_c$ than $\omega_p^c/\sqrt{\varepsilon_\infty^c}$ of the coherent free carriers for x>0.07. Further studies are necessary to understand the physical origin of the almost dissipationless $c$-axis transport of the free carriers and the pairing mechanism responsible for their eventual condensation into a global high-temperature superconductor.

We are indebted to Y. S. Song for his technical assistance and W. X. Ti for his work in crystal growth. The work in Beijing is supported by the National Natural Science Foundation of China (project 10174090) and the Ministry of Science and Technology of China (project G1999064601). The work conducted in Houston is supported by the State of Texas through the Texas Center for Superconductivity and Advanced Materials.